\documentclass[12pt]{iopart}
\usepackage{graphicx}
\usepackage{bm}
\usepackage{iopams} 

\begin{document}

\article{Fast Track Communication}{Tracking the energies of
one-dimensional subband edges in quantum point contacts using dc
conductance measurements}

\author{A.P. Micolich}
\address{School of Physics, University of New South Wales, Sydney NSW 2052, Australia}
\ead{adam.micolich@nanoelectronics.physics.unsw.edu.au}

\author{U. Z\"{u}licke}
\address{School of Chemical and Physical Sciences and MacDiarmid Institute for Advanced Materials and Nanotechnology, Victoria University of Wellington, Wellington 6140, New Zealand}
\ead{Uli.Zuelicke@vuw.ac.nz}

\submitto{\JPCM}
\date{\today}

\begin{abstract}

The semiconductor quantum point contact has long been a focal point
for studies of one-dimensional electron transport. Their electrical
properties are typically studied using ac conductance methods, but
recent work has shown that the dc conductance can be used to obtain
additional information, with a density-dependent Land\'{e} effective
$g$-factor recently reported [T.-M. Chen {\it et al}, Phys. Rev. B
{\bf 79}, 081301 (2009)]. We discuss previous dc conductance
measurements of quantum point contacts, demonstrating how valuable
additional information can be extracted from the data. We provide a
comprehensive and general framework for dc conductance measurements
that provides a path to improving the accuracy of existing data and
obtaining useful additional data. A key aspect is that dc
conductance measurements can be used to map the energy of the 1D
subband edges directly, giving new insight into the physics that
takes place as the spin-split 1D subbands populate. Through a
re-analysis of the data obtained by Chen {\it et al}, we obtain two
findings. The first is that the $2\downarrow$ subband edge closely
tracks the source chemical potential when it first begins populating
before dropping more rapidly in energy. The second is that the
$2\uparrow$ subband populates more rapidly as the subband edge
approaches the drain potential. This second finding suggests that
the spin-gap may stop opening, or even begin to close again, as the
$2\uparrow$ subband continues populating, consistent with recent
theoretical calculations and experimental studies.

\end{abstract}
\maketitle

\section{Introduction}

The Quantum Point Contact (QPC) is a major landmark in the study of
the electronic properties of nanoscale devices~\cite{BerggrenPW02}.
Advanced semiconductor production techniques such as molecular beam
epitaxy~\cite{ChoAPL71} allow AlGaAs/GaAs heterostructures to be
grown with monolayer precision. Such structures can support a buried
two-dimensional electron gas (2DEG) that can be patterned
electrostatically using metal `gates' on the heterostructure
surface. A QPC is typically defined by using a
split-gate~\cite{ThorntonPRL86}, a strip of metal with a $\sim
1~\mu$m gap in the middle, which separates the 2DEG into source and
drain reservoirs either side of an aperture of width comparable to
the electron Fermi wavelength ($\sim 50$~nm). The width of the
aperture can be tuned by adjusting the voltage $V_{g}$ applied to
the gates, while the only requirement on length is that it is less
than the elastic mean free path ($\sim 1-10~\mu$m) so that transport
through the aperture is ballistic. At low temperature $T \lesssim
5$~K, the linear conductance $G$ reduces in quantized steps of
$G_{0} = 2e^{2}/h$ as the QPC is narrowed. This is due to
depopulation of the 1D subbands within the QPC as they rise up above
the Fermi energy in the reservoirs~\cite{vanWeesPRL88, WharamJPC88},
providing a striking demonstration of the importance of quantum
effects in the operation of nanoscale devices.

A curious non-quantized plateau-like feature observed at $G = 0.7
G_{0}$, first reported by Thomas {\it et al} in
1996~\cite{ThomasPRL96}, has drawn significant attention. The origin
of this effect is still a matter of debate -- while it is widely
accepted as a many-body phenomenon, numerous microscopic mechanisms
have been proposed including the presence of a static
spin-polarization due to the exchange interaction~\cite{ThomasPRL96,
WangPRB98} and a manifestation of the Kondo effect~\cite{MeirPRL02,
CronenwettPRL02}, amongst others. There have also been several
phenomenological models proposed based on the opening of an energy
gap between the spin-up and spin-down components of the 1D subbands.
These have successfully reproduced much of the essential behaviour
observed experimentally for the $0.7$ plateau~\cite{KristensenPRB00,
BruusPhysE01, ReillyPRL02, ReillyPRB05, ReillyPhysE06}. Recent
experiments by the Cambridge group have shed interesting new light
on this problem. Studies of analogous non-quantized plateaus at $G >
G_{0}$ called `$0.7$ analogs' by Graham {\it et al} suggest that the
spin-down subbands drop rapidly in energy upon
population~\cite{GrahamPRB05} while the spin-up subband edges pin at
the chemical potential~\cite{GrahamPRB07}, in general support of the
spin-gap models.

Recently, Chen {\it et al} revealed that new information could be
obtained by combining measurements of the dc conductance with the
more commonly measured ac conductance~\cite{ChenAPL08, ChenPRB09,
ChenNL10}. Chen {\it et al} show additional evidence for differences
in the population rates of spin-up and spin-down subbands based on
plotting the ac and dc transconductance $dG/dV_{g}$ as a colour-map
against gate voltage $V_{g}$ and the dc source-drain bias
$V_{sd}$~\cite{ChenPRB09}. Additionally, they report measurements of
the effective Land\'{e} $g$-factor $g^{*}$ versus $V_{g}$, which
exhibit a sawtooth appearance indicative of increasing $g$-factor
with increasing electron density. This result seems counterintuitive
considered alongside numerous preceding experiments showing instead
that $g^{*}$ obtained from ac conductance measurements increases as
the density is reduced in a QPC~\cite{ThomasPRL96, PatelPRB91,
DaneshvarPRB97, DanneauPRL06, MartinAPL08, MartinPRB10}.

The purpose of this paper is to take a further look at dc
conductance measurements of QPCs and the information that can be
gleaned from them. We will begin in Section 2 with a brief
discussion of the dc conductance technique. In Section 3 we point
out where the $g$-factor data presented by Chen {\it et al} has
reduced accuracy, and suggest additional measurements that can
provide both improved precision and new information that may have
implications for the future interpretation of $g$-factor
measurements obtained from the dc conductance. Finally, in Section 4
we present a re-analysis of the data presented by Chen {\it et al}
in Ref.~\cite{ChenPRB09} showing that the dc conductance can be used
instead to map the evolution of the 1D subbands in energy with gate
voltage. This re-analysis shows very clearly that the spin-down
subbands drop rapidly in energy when they populate, in agreement
with earlier data by Graham {\it et al}~\cite{GrahamPRB05}, and that
the spin-up subbands populate much more slowly (a key conclusion of
the paper by Chen {\it et al}~\cite{ChenPRB09}), at least initially.
We find that the population rate of the spin-up subband increases as
$V_{g}$ is made more positive, becoming comparable to that of the
spin-down subband as the spin-up subband edge approaches the drain
potential. An interesting additional feature is observed: the
spin-down subband edge appears to briefly track close to the source
chemical potential, indicating delayed initial population for the
spin-down subband also. This trend is observed for both the
$2\downarrow$ and $3\downarrow$ subbands, albeit more strongly for
the former. Our re-analysis and the limited data available in
Ref.~\cite{ChenPRB09} demonstrate that more extensive measurements
using this approach are warranted.

\section{How the ac and dc conductances differ: A brief primer on dc conductance measurements}

Experimental studies of electron transport in QPCs have
traditionally relied on measurements of the ac conductance $G_{ac} =
I^{ac}_{sd}/V^{ac}_{sd}$, typically obtained by applying a
$10-100~\mu$V ac bias $V^{ac}_{sd}$ at a frequency of $\sim
5-300$~Hz to the source, and measuring the resulting current
$I^{ac}_{sd}$ at the drain using a lock-in amplifier. These
measurements can be performed with the addition of a dc bias
$V^{dc}_{sd}$ to the ac bias used to obtain $G_{ac}$ using a simple
adder circuit. The dc bias separates the source $\mu_{s}$ and drain
$\mu_{d}$ chemical potentials in energy by $\mu_{s} - \mu_{d} =
eV^{dc}_{sd}$~\footnote{Note that a positive bias applied to one
reservoir actually lowers it in energy with respect to the other.
Hence there is an implicit sign reversal assumed here, such that the
drain is held fixed at electrical ground and a positive
$V^{dc}_{sd}$ raises $\mu_{s}$ above $\mu_{d}$ rather than lowering
it}, allowing spectroscopic measurements of the 1D subband edges to
be performed~\cite{KristensenPRB00, PatelPRB91A} (see Section 3).
Very recently, measurements of the dc conductance $G_{dc} =
I^{dc}_{sd}/V^{dc}_{sd}$ have been used to gain further insight into
transport in QPCs~\cite{ChenAPL08, ChenPRB09, ChenNL10}. The dc
conductance can be measured simultaneously with the ac conductance
by passing the output current from the device into a preamplifier to
convert it to a voltage, and on into a lock-in amplifier and dc
multimeter in a parallel circuit to ground~\cite{pc}.

The physical difference between the two conductivities is
significant as they provide very different information about changes
in the energy of the 1D subband edges. The ac conductance is a
differential conductance, representing the gradient of the full dc
$I$-$V$ curve for the QPC over a pair of narrow bias windows of
width $V^{ac}_{sd}$, one centered on $\mu_{s}$ and the other on
$\mu_{d}$. Because $V^{ac}_{sd}$ is usually kept small to minimise
heating/broadening, $G_{ac}$ is only sensitive to a 1D subband edge
passing through a chemical potential, giving a quantized jump in
$G_{ac}$ of $G_{0}$ for a spin-degenerate subband in the zero dc
bias limit where $\mu_{s} = \mu_{d} = \mu$, and a step of $0.25
G_{0}$ when a spin-polarized subband edge passes through $\mu_{s}$
or $\mu_{d}$ if $\mu_{s} - \mu_{d} > eV^{ac}_{sd}$. In contrast, the
dc conductance is sensitive to subband edge motion through the
entire window between $\mu_{s}$ and $\mu_{d}$. Consider a 1D subband
that starts above $\mu_{s}$ which in turn is $eV^{dc}_{sd} >>
eV^{ac}_{sd}$ above $\mu_{d}$. As the 1D subband falls in energy,
$G_{dc}$ remains constant until the subband edge crosses $\mu_{s}$.
The dc conductance then increases gradually as the subband edge
lowers through the bias window, ultimately reaching $\mu_{d}$ where
$G_{dc}$ once again becomes constant, having increased by $G_{0}$
for a spin-degenerate subband and $0.5 G_{0}$ for a spin-polarized
subband. The crucial aspect for this study is that $G_{dc}$ provides
information about the location and rate of movement of a 1D subband
edge as a function of $V_{g}$ whenever it is located between
$\mu_{s}$ and $\mu_{d}$, information that cannot be attained from ac
conductance measurements alone~\cite{ChenAPL08}.

\section{Measurements of the $g$-factor using dc conductance}

Measurements of $G_{ac}$ and $G_{dc}$ versus $V_{g}$ as a function
of $V^{dc}_{sd}$ can be used as a spectroscopic tool for studying
the physics of the 1D subbands~\cite{ChenPRB09, ChenNL10}. This is
often achieved using a greyscale plot or colour-map of the
transconductance $dG_{ac}/dV_{g}$ or $dG_{dc}/dV_{g}$ versus $V_{g}$
($x$-axis) and $V^{dc}_{sd}$ ($y$-axis) -- examples of such
colour-maps for ac and dc transconductance appear in Figs.~3(a) and
(b) of Ref.~\cite{ChenPRB09}, respectively. Regions of low and high
$dG/dV_{g}$ (dark and bright in Fig.~3 of Ref.~\cite{ChenPRB09})
indicate conductance plateaus and rises in conductance between
plateaus, respectively. In the ac case, if we start at $V_{sd} = 0$
and increase the dc bias, the high $dG/dV_{g}$ regions evolve into
V-shaped structures. The left- and right-moving branches correspond
to a given 1D subband edge coinciding with $\mu_{s}$ and $\mu_{d}$,
respectively. The area inside the V-shaped structure has
$dG_{ac}/dV_{g} \approxeq 0$ as $G_{ac}$ is fixed when the subband
edge is not within $V^{ac}_{sd}$ of $\mu_{s}$ or $\mu_{d}$. In
contrast, for a dc transconductance colour-map the V-shaped region
is `filled', with $dG_{dc}/dV_{g}$ indicating the rate at which the
subband edge moves in energy between $\mu_{s}$ and $\mu_{d}$.

\begin{figure}
\includegraphics[width=16cm]{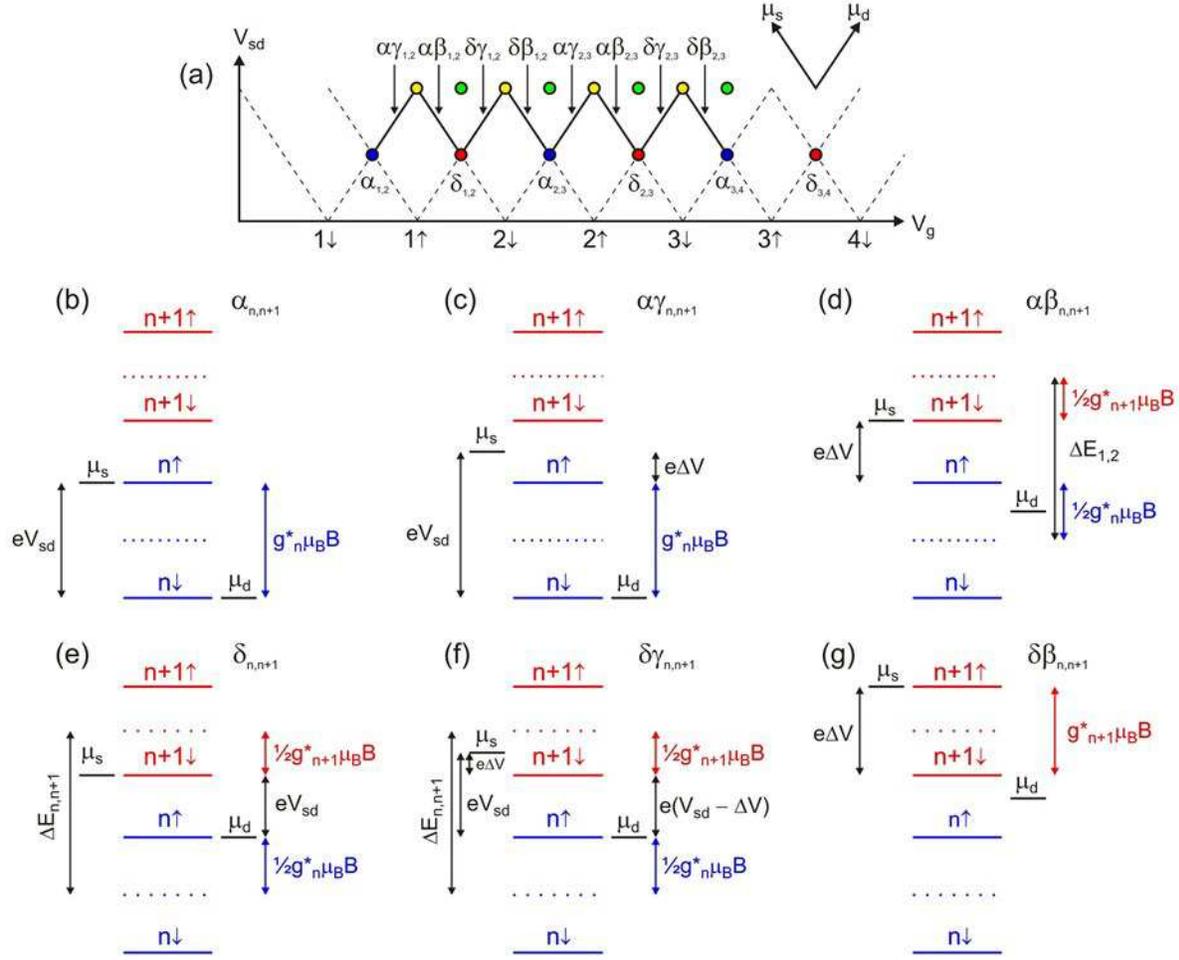}
\caption{(a) Schematic of an idealised transconductance map plotted
against dc source-drain bias $V_{sd}$ ($y$-axis) and gate voltage
$V_{g}$ ($x$-axis) for a large in-plane magnetic field such that the
spin-degeneracy of the 1D subbands is broken. The diagram contains a
number of diagonal dashed lines forming V-shaped structures with a
vertex on the $V_{g}$ axis. These vertices correspond to the $V_{g}$
at which a subband with index $n$ and spin up ($\uparrow$) or down
($\downarrow$) intercepts the chemical potential $\mu$. The left
(right) branches of a given V-shaped structure correspond to that
subband coinciding with the source $\mu_{s}$ (drain $\mu_{d}$)
potentials. The blue and red circles indicate points where $\mu_{s}$
and $\mu_{d}$ coincide with adjacent 1D subband edges, these being
the $n\uparrow$ and $n\downarrow$ ($n\uparrow$ and $n+1\downarrow$)
subband edges for the blue (red) circles. The 1D subband spacing can
be obtained at these points following Patel {\it et
al}~\cite{PatelPRB91A}. The yellow circles indicate points where
$\mu_{s}$ and $\mu_{d}$ coincide with subband edges having the same
spin but subband index differing by 1. The separations $\Delta E
_{n\uparrow,n+1\uparrow}$ and $\Delta E
_{n\downarrow,n+1\downarrow}$ are obtained at these points, which
play an essential role in the method used by Chen {\it et al} to
obtain $g^{*}$. The green circles indicate crossing points for
spin-degenerate subband edges, where the separations $\Delta
E_{n,n+1}$ would be obtained at $B_{\parallel} = 0$. The solid
zig-zag line indicates the path taken in obtaining $g^{*}$ data,
with the results in Ref.~\cite{ChenPRB09} obtained from short
segments of such a path. (b-g) Subband edge energy diagrams
corresponding to the six configurations in (a). The quantity
$e\Delta V$ is the energy separation between $\mu_{s}$ and the next
lowest subband edge, and is obtained using $G_{dc} = (n + \Delta
V/V_{sd})G_{0}$. Other quantities are defined in the text. These are
(b) $\alpha_{n,n+1}$ blue circles, (c) $\alpha\gamma_{n,n+1}$
positive-gradient part of zig-zag, (d) $\alpha\beta_{n,n+1}$
negative-gradient part of zig-zag, (e) $\delta_{n,n+1}$ red circles,
(f) $\delta\gamma_{n,n+1}$ positive-gradient part of zig-zag and (g)
$\delta\beta_{n,n+1}$ negative-gradient part of zig-zag.}
\end{figure}

To facilitate further discussion, we refer to the schematic
transconductance map shown in Fig.~1(a), where the numbers
correspond to the 1D subband index $n = 1,2,3,...$, and $\uparrow$
and $\downarrow$ to spin up and down respectively. For direct
comparison with Ref.~\cite{ChenPRB09}, we consider the case where a
strong in-plane magnetic field $B_{\parallel}$ is applied, breaking
the spin-degeneracy of the 1D subbands. The left- and right-sloping
dashed diagonal lines in Fig.~1(a) indicate the gate voltage
settings where a given 1D subband edge coincides with $\mu_{s}$ and
$\mu_{d}$, respectively, as the dc bias $V_{sd}$ is increased. Note
that although these are presented as straight lines in our
schematic, in reality they curve slightly due to the gate voltage
dependent subband spacing caused by the self-consistent
electrostatic potential of the QPC~\cite{ButtikerPRB90}. The
$g$-factor measurements in Ref.~\cite{ChenPRB09} are obtained by
following a very specific zig-zag path through the transconductance
plot, indicated by the thick black line in Fig.~1(a) (c.f. Fig.~3(a)
of Ref.~\cite{ChenPRB09}). The path has six general `configurations'
from which an estimate of $g^{*}$ can be obtained. These
configurations repeat cyclically moving right or left along the
zig-zag path, and correspond to the six energy diagrams shown in
Fig.~1(b-g) where we plot the positions of subband edges $n
\uparrow$, $n \downarrow$, $n+1 \uparrow$ and $n+1 \downarrow$
relative to $\mu_{s}$ and $\mu_{d}$. These are labelled
$\alpha_{n,n+1}$, $\alpha\gamma_{n,n+1}$, $\alpha\beta_{n,n+1}$,
$\delta_{n,n+1}$, $\delta\gamma_{n,n+1}$ and $\delta\beta_{n,n+1}$
to correspond as directly as possible to the three configurations
$\alpha$, $\beta$ and $\gamma$ identified for the 1st subband data
in Fig.~4 of Ref.~\cite{ChenPRB09}, and provide a more general
framework for measurements of $g^{*}$.

\subsection{General framework for extracting $g^{*}$ from high-field
source-drain bias spectroscopy}

At the low $V_{sd}$ vertices, highlighted by six circles that
alternate between blue and red in Fig.~1(a), 1D subband edges
adjacent in energy align with $\mu_{s}$ and $\mu_{d}$, allowing
their energy separation to be read directly as $eV_{sd}$. These six
points only require measurement of the ac conductance, and represent
the method established by Patel {\it et al}~\cite{PatelPRB91A} and
used previously~\cite{ThomasPRL96, PatelPRB91, DaneshvarPRB97,
DanneauPRL06, SchapersAPL07, MartinAPL08, MartinPRB10} to determine
the $g$-factor of the 1D subbands. The blue circles correspond to
the $\alpha_{n,n+1}$ configuration (see Fig.~1(b) and 2(a)), where
the $n\uparrow$ and $n\downarrow$ subband edges align with $\mu_{s}$
and $\mu_{d}$, respectively, and a pure, single subband $g^{*}$
value can be directly measured. The red circles correspond to the
$\delta_{n,n+1}$ configuration (see Fig.~1(e)), here direct
measurement using $eV_{sd}$ at the subband crossing gives instead an
average of the $g$-factors for two adjacent subbands $(g^{*}_{n} +
g^{*}_{n+1})/2$~\cite{DanneauPRL06} unless additional information is
known about the relative locations of various 1D subband edges, as
applied by Chen {\it et al}~\cite{ChenPRB09} (see Section 3.2).

Starting at $\alpha_{n,n+1}$ or $\delta_{n,n+1}$ has $\mu_{s}$ and
$\mu_{d}$ held at adjacent subband edges. Following the zig-zag in
Fig.~1(a) involves first holding the lower 1D subband edge at
$\mu_{d}$ while raising $\mu_{s}$ up towards the next highest
subband along the positive-gradient diagonal (see
$\alpha\gamma_{n,n+1}$ in Fig.~1(c) and $\delta\gamma_{n,n+1}$ in
Fig.~1(f)), and then holding $\mu_{s}$ at that subband edge while
$\mu_{d}$ rises up to the subband directly below along the negative
gradient diagonal (see $\alpha\beta_{n,n+1}$ in Fig.~1(d) and
$\delta\beta_{n,n+1}$ in Fig.~1(g)). Note that moving right in
Fig.~1(a) corresponds to all of the 1D subbands moving downwards
together in energy. It is the vertical motion associated with the
zig-zag that allows one of $\mu_{s}$ or $\mu_{d}$ to track one
subband edge while keeping the adjacent 1D subband edge in the bias
window to allow continuous measurement of $g^{*}$ via $G_{dc}$
(e.g., see the horizontal purple bars in Fig.~3). For all positions
on the $\alpha\gamma_{n,n+1}$, $\alpha\beta_{n,n+1}$,
$\delta\gamma_{n,n+1}$ and $\delta\beta_{n,n+1}$ branches, the dc
conductance $G_{dc} = (n + \Delta V/V_{sd})G_{0}$ for
spin-degenerate 1D subbands and $G_{dc} = (n + \Delta
V/V_{sd})e^{2}/h$ for spin-polarised 1D subbands is needed to
establish the energy separation $e\Delta V$ between the $\mu_{s}$
and the edge of the subband edge sitting below it in the bias window
in order to obtain information about $g^{*}$.

Finally, at the upper vertices of the zig-zag (yellow circles in
Fig.~1(a)), $\mu_{s}$ and $\mu_{d}$ span two subband gaps, and these
points allow direct local measurements of the energy separation
between identical spin branches of adjacent index subbands (i.e.,
$\Delta E_{n\uparrow,n+1\uparrow} = E_{n+1\uparrow} - E_{n\uparrow}$
or $\Delta E_{n\downarrow,n+1\downarrow}$). Although Chen {\it et
al} do not mention this explicitly, a measurement of this gap is
vital to their method for obtaining single-subband $g^{*}$ values.
It is the cause of significant problems for the accuracy of these
$g^{*}$ values, as we now discuss.

\begin{figure}
\includegraphics[width=10cm]{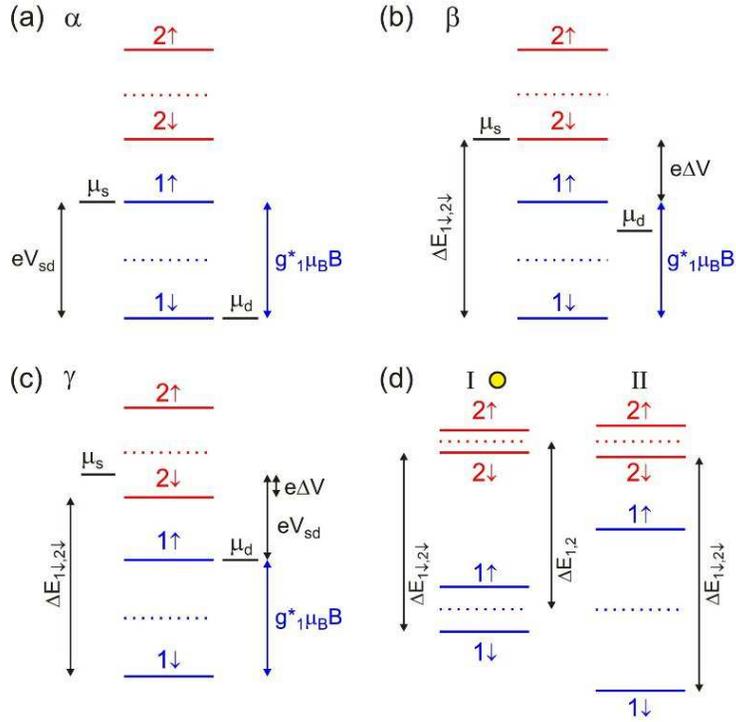}
\caption{(a - c) Subband edge energy diagrams illustrating the three
configurations $\alpha$, $\beta$ and $\gamma$ used in
Ref.~\cite{ChenPRB09}. (d) Schematics illustrating why $\Delta
E_{1\downarrow,2\downarrow}$ cannot be safely assumed to be
independent of $V_{g}$. For sake of argument, Schematic I is chosen
to coincide with the left-most yellow circle in Fig.~1(a), where the
1st and 2nd subband $g$-factors $g^{*}_{1}$ and $g^{*}_{2}$ are
small in comparison to the spin-degenerate subband spacing $\Delta
E_{1,2}$. The separation $\Delta E_{1\downarrow,2\downarrow}$ used
by Chen {\it et al} to obtain $g^{*}$ depends on $g^{*}_{1}$,
$g^{*}_{2}$ and $\Delta E_{1,2}$ as given in Eq.~(1). As $V_{g}$ is
made more positive (i.e., moving right in Fig.~1(a) here and Fig.~4
of Ref.~\cite{ChenPRB09}), three changes occur: $g^{*}_{1}$
increases dramatically, $\Delta E_{1,2}$ decreases slightly due to
weakening 1D confinement, and $g^{*}_{2}$ may also increase, albeit
to a much lesser extent than $g^{*}_{1}$. Ultimately, this makes it
impossible for $\Delta E_{1\downarrow,2\downarrow}$ to remain
constant, introducing significant systematic error into the $g^{*}$
values presented by Chen {\it et al}, as discussed in the text.}
\end{figure}

\subsection{A review of $g^{*}$ data obtained from dc conductance measurements}

The $g^{*}$ data in Ref.~\cite{ChenPRB09} is obtained from five
short segments from a zig-zag path such as that in Fig.~1(a). We
start this discussion by pointing out how this data, and the three
configurations $\alpha$, $\beta$ and $\gamma$ they identify for the
first subband in Fig.~4 of Ref.~\cite{ChenPRB09}, correspond to the
more general framework discussed above and presented in Fig.~1.

The $\alpha$ configuration in Ref.~\cite{ChenPRB09} shown in
Fig.~2(a) corresponds exactly to $\alpha_{1,2}$ (see Fig.~1(b)). In
the data presented by Chen {\it et al} this configuration provides
the lowest $g^{*}$ value at most negative $V_{g}$ for each subband
(first, third and fifth open symbols in Fig.~4 of
Ref.~\cite{ChenPRB09}). The next stretch of values, solid symbols
denoted $\beta$ in Ref.~\cite{ChenPRB09}, correspond to the energy
diagram in Fig.~2(b) and to the negative-gradient
$\alpha\beta_{1,2}$ branch in Fig.~1(a) (i.e., the
$\alpha\gamma_{1,2}$ branch is skipped in the measurements in
Ref.~\cite{ChenPRB09}). The open symbol between the $\beta$ and
$\gamma$ branches corresponds to $\delta_{1,2}$ in the nomenclature
used in Fig.~1, and is the situation intermediate to Figs.~2(b) and
(c) where $\mu_{s}$ ($\mu_{d}$) coincides with $2\downarrow$
($1\uparrow$). Finally the $\gamma$ branch in Fig.~4 of
Ref.~\cite{ChenPRB09} corresponds to the positive-gradient
$\delta\gamma_{1,2}$ branch in Fig.~1 and the energy diagram in
Fig.~2(c)\footnote{A careful analysis of Figs.~3 and 4 in
Ref.~\cite{ChenPRB09} reveals that the fifth point on the $\gamma$
branch for the 1st subband lies beyond the crossing point for the
$1\uparrow$ and $2\uparrow$ subbands, while the 4th sits very close
to this crossing and possibly just to the right of it. Certainly in
the case of the 5th point, and perhaps the 4th, there is more than
one 1D subband edge within the bias window, and these data points
should thus be considered with caution.}.

\subsubsection{Accuracy of the $g^{*}$ values}

It is important to note that with the exception of the one point
marked $\alpha$, the extraction of all of the first subband $g^{*}$
values in Fig.~4 of Ref.~\cite{ChenPRB09} relies on the implicit
assumption that the subband separation $\Delta
E_{1\downarrow,2\downarrow}$ is constant as a function of $V_{g}$.
Significant caution needs to be exercised in making this assumption,
because as Fig.~2(d) highlights, $\Delta
E_{1\downarrow,2\downarrow}$ is not constant.

Consider Schematic I in Fig.~2(d); for the sake of argument, let us
assume that this corresponds to the left-most yellow circle in
Fig.~1(a), which is the only gate voltage at which a precise
measured value for $\Delta E_{1\downarrow,2\downarrow}$ can be
obtained. The dashed horizontal lines indicate the spin-degenerate
edges of the first and second subbands, their separation $\Delta
E_{1,2}$ is set by the QPC confinement potential. With a magnetic
field $B_{\parallel}$ applied, the spin degeneracy is broken, and
the $1\downarrow$ and $1\uparrow$ subband edges separate in energy
by $g^{*}_{1}\mu_{B}B_{\parallel}$ while the $2\downarrow$ and
$2\uparrow$ subband edges separate by
$g^{*}_{2}\mu_{B}B_{\parallel}$. Note that $g^{*}_{1}$ does not
necessarily equal $g^{*}_{2}$. As Schematic I shows, this results
in:

\begin{equation}
\Delta E_{1\downarrow,2\downarrow} = \Delta E_{1,2} +
g^{*}_{1}\mu_{B}B_{\parallel} - g^{*}_{2}\mu_{B}B_{\parallel}
\end{equation}

\noindent If we accept the premise that $g^{*}$ is $V_{g}$ dependent
in the manner suggested by Fig.~4 in Ref.~\cite{ChenPRB09}, then
making $V_{g}$ more positive would bring us to the scenario in
Schematic II in Fig.~2(d), where two changes will have occurred.
First and foremost, $g^{*}_{1}$ will have increased significantly,
as we discuss below. Second, as $V_{g}$ becomes more positive the
confinement is weakened, reducing the separation $\Delta E_{1,2}$.
There may also be an increase in $g^{*}_{2}$, although this should
be a much smaller contribution. Thus if we consider Eq.~(1), it is
clear that $\Delta E_{1\downarrow,2\downarrow}$ cannot possibly be
constant as a function of $V_{g}$.

Some simple estimates confirm the significance of this issue. The
separation $\Delta E_{1\downarrow,2\downarrow}$ can be directly and
accurately measured at $V_{g} = -5.4785$~V, and using the data in
Fig.~3(a) of Ref.~\cite{ChenPRB09}, we obtain $\Delta
E_{1\downarrow,2\downarrow} = 3.33$~meV. If one considers the data
in Fig.~4 of Ref.~\cite{ChenPRB09}, the separation between
$1\downarrow$ and $1\uparrow$ (i.e., $g^{*}_{1}\mu_{B}B$) increases
by a factor of over $3.5$ from $0.8$~meV at $V_{g} = -5.5431$~V
($\alpha$ point) to $2.87$~meV at $-5.37$~V (right-most point of
$\gamma$ branch). This corresponds to the full range over which
$\Delta E_{1\downarrow,2\downarrow}$ is assumed constant in
determining $g^{*}$. Clearly the effect of the change in $g^{*}_{1}$
on $\Delta E_{1\downarrow,2\downarrow}$ is impossible to neglect,
and the gate voltage dependence of $\Delta E_{1,2}$ will exacerbate
this.

The $\Delta E_{n\uparrow,n+1\uparrow}$ or $\Delta
E_{n\downarrow,n+1\downarrow}$ obtained at an upper subband vertex
is a reasonable approximation on the branches running down either
side from the corresponding vertex (i.e., $\alpha\gamma_{n,n+1}$ and
$\alpha\beta_{n,n+1}$ for $\Delta E_{n\downarrow,n+1\downarrow}$,
and $\delta\gamma_{n,n+1}$ and $\delta\beta_{n,n+1}$ for $\Delta
E_{n\uparrow,n+1\uparrow}$). However, by the final $\gamma$ point in
Fig.~4 of Ref.~\cite{ChenPRB09}, which is obtained at the top of the
$\delta\gamma_{1,2}$ branch using a $\Delta
E_{1\downarrow,2\downarrow}$ estimate obtained at the vertex of the
$\alpha\gamma_{1,2}$ and $\alpha\beta_{1,2}$ branches, the $g^{*}$
values become inaccurate. The accuracy may be improved by making
full use of available subband spacing measurements. For example, at
$B_{\parallel} = 0$ where the $\uparrow$ and $\downarrow$ branches
are degenerate, the subband spacing $\Delta E_{1,2}$ can be
measured. The corresponding points are shown as green circles in
Fig.~1(a), and the diagrams in Fig.~1(b-g) indicate how this can be
used to obtain additional $g^{*}$ estimates. Furthermore, the
$\Delta E_{n\uparrow,n+1\uparrow}$ estimates can be used to obtain
further measurements beyond those presented in
Ref.~\cite{ChenPRB09}, as we discuss below. Putting all these
measurements together and carefully considering the range of
validity for each subband spacing, a more accurate picture can be
obtained.

\subsubsection{Additional data from $\alpha\gamma_{n,n+1}$ and $\delta\beta_{n,n+1}$ branches}

Considering the data in Ref.~\cite{ChenPRB09} alongside Fig.~1 it is
clear that while Chen {\it et al} have made an important
contribution, the opportunity exists for a substantial amount of
additional data to be obtained with their approach. This may point
to behaviour more complex than the linear trend shown in Fig.~4 of
Ref.~\cite{ChenPRB09}. The missing $\alpha\gamma_{1,2}$ branch for
the first subband would provide information about what happens as
the $1\uparrow$ subband begins to populate, and it would be
interesting to compare this with the behaviour of the $\gamma$
branch in Fig.~4 of Ref.~\cite{ChenPRB09}. The $\gamma$ branch
($\delta\gamma_{1,2}$ in our nomenclature) corresponds to the
initial population of $2\downarrow$, and the first point appears to
deviate from the linear trend of the rest of the branch (we return
to this in Section 4). It is interesting to note that, in contrast
to the first subband data, the $\alpha\gamma_{2,3}$ branch is
measured for the second subband in Ref.~\cite{ChenPRB09}, and the
non-monoticity (albeit with only three data points) may suggest that
interesting behaviour occurs as $2\uparrow$ populates.

The missing $\delta\beta$ branch for both the 1st and 2nd subbands
would also be interesting, particularly combined with the additional
data that could be obtained using the $\Delta
E_{n\uparrow,n+1\uparrow}$ measurements. In Ref.~\cite{ChenPRB09},
the $\Delta E_{1\downarrow,2\downarrow}$ value obtained at the
intersection of the $\alpha\gamma$ and $\alpha\beta$ branches is
used to derive $g^{*}_{1}$ along the $\gamma$ branch (i.e.,
$\delta\gamma$ branch in Fig.~1(a)). However, $\Delta
E_{1\downarrow,2\downarrow}$, which is obtained at the
$\delta\gamma$/$\delta\beta$ intersection and is thus more accurate
here, can be used to measure $g^{*}_{2}$ along this same $\gamma$
branch as well. With $g^{*}_{2}$ measured along both $\delta\gamma$
and $\delta\beta$ branches, it would be possible to establish
precisely how the gap between $2\downarrow$ and $2\uparrow$ evolves
as the $2\downarrow$ subband populates. It may also be possible to
adapt this process to the left of the first $\alpha$ point in Fig.~4
of Ref~\cite{ChenPRB09} to obtain useful information regarding the
population of the $1\downarrow$ subband. However, since there is no
subband separation information to the left of where $\Delta
E_{1\downarrow,2\downarrow}$ is obtained, this data may be
qualitative at best.

\subsubsection{Interpretation of the $g$-factor data}

The $g^{*}$ data presented in Ref.~\cite{ChenPRB09} are a surprising
contrast to the previously accepted trend for $g^{*}$ to gradually
and monotonically increase as the subband index $n$ is
reduced~\cite{ThomasPRL96, PatelPRB91, DaneshvarPRB97, DanneauPRL06,
MartinAPL08, MartinPRB10}. Chen {\it et al} note the remarkable
similarity between these data and both the oscillatory $g^{*}$ in
quantum Hall systems as consecutive Landau levels are
filled~\cite{AndoJPSJ74} and the theoretical prediction by Wang and
Berggren (see Fig.~2 of Ref~\cite{WangPRB96}). The latter is of
particular interest, as a closer inspection reveals that the data
and this prediction actually disagree. Firstly, we consider exactly
what Wang and Berggren predict.

Figure~2 of Wang and Berggren's paper~\cite{WangPRB96} shows
$g^{*}$, calculated using density functional theory, for the lowest
spin-split subband as the 1D density within the QPC is increased
such that the second, third and fourth 1D subbands populate. This
calculated $g^{*}_{1}$ rises at the point where a spin-degenerate
subband begins to populate as exchange effects lead to a spontaneous
spin polarization and a finite spin-gap (i.e., separation between
spin-up and spin-down components of that particular subband). This
gap collapses once the opposite spin subband edge drops below the
Fermi energy, driving $g^{*}$ back towards zero. The result is an
`undulating' $g^{*}_{1}$ versus $n_{1D}$ where the undulations get
smaller as successively higher subbands fill. The behaviour Wang and
Berggren calculate in Fig.~2 of Ref.~\cite{WangPRB96} cannot be
measured directly in the conductance. Looking at Fig.~1, the
$\delta\gamma_{1,2}$ branch (i.e., $\gamma$ in Fig.~4 of
Ref.~\cite{ChenPRB09}) is the last point where any direct
information about $g^{*}_{1}$ can be obtained because from
$\delta\beta_{1,2}$ onwards the edges of both $1\downarrow$ and
$1\uparrow$ move below $\mu_{d}$ and away from the bias window.

However, from Fig.~1(a/b) of Ref.~\cite{WangPRB96} it is clear that
whenever an exchange induced spin-gap occurs in the first subband,
it also occurs in the higher subbands. Thus in a measurement such as
that presented by Chen {\it et al} where the focus needs to shift
from one subband to the next as they fall below the dc bias window,
the calculations by Wang and Berggren would still predict an
oscillatory behaviour of $g^{*}$ (providing that there are no
obscuring artifacts due to the change in the particular subband
being measured). This can be determined by mapping the locations of
the rises and falls in $g^{*}$ to where the subband edges pass into
and out of the bias window, and this is where we find an important
discrepancy.

Returning to Wang and Berggren's calculation, $g^{*}_{1}$ only rises
until the $1\uparrow$ subband falls below the chemical potential; it
collapses back to almost zero thereafter. This would occur at point
$\delta_{1,2}$, and so the continued rise in $g^{*}_{1}$ along the
$\gamma$ branch in Fig.~4 of Ref.~\cite{ChenPRB09} contradicts the
calculated behaviour in Ref~\cite{WangPRB96}. Note that the apparent
precipitous drop in $g^{*}$ at the end of the $\gamma$ branch
actually reflects a change in what is measured from $g^{*}_{1}$ to
$g^{*}_{2}$, which is significantly smaller in magnitude. Indeed,
the data in Fig.~4 of Ref.~\cite{ChenPRB09} is more consistent with
the density-dependent spin-gap model developed by Reilly {\it et
al}~\cite{ReillyPRL02, ReillyPRB05, ReillyPhysE06}, and obtaining
measurements of $g^{*}_{2}$ over the $\delta\gamma_{1,2}$ and
$\delta\beta_{1,2}$ branches either side of the
$1\uparrow$/$2\uparrow$ subband crossing would be particularly
enlightening in this regard.

\section{Tracking the 1D subbands}

We finish by considering an interesting question -- why measure the
$g$-factor at all? In obtaining the $g$-factor by the method used in
Ref.~\cite{ChenPRB09}, we take rather precise information about the
location of two adjacent subband edges, one held at a chemical
potential and the other measured relative to it using the dc
conductance, and combine it with comparatively imprecise information
about the energy separation between other subband edges. Why not
confine our attention to the subband measurements since this should
provide more precise and useful information.

\begin{figure}
\includegraphics[width=12cm]{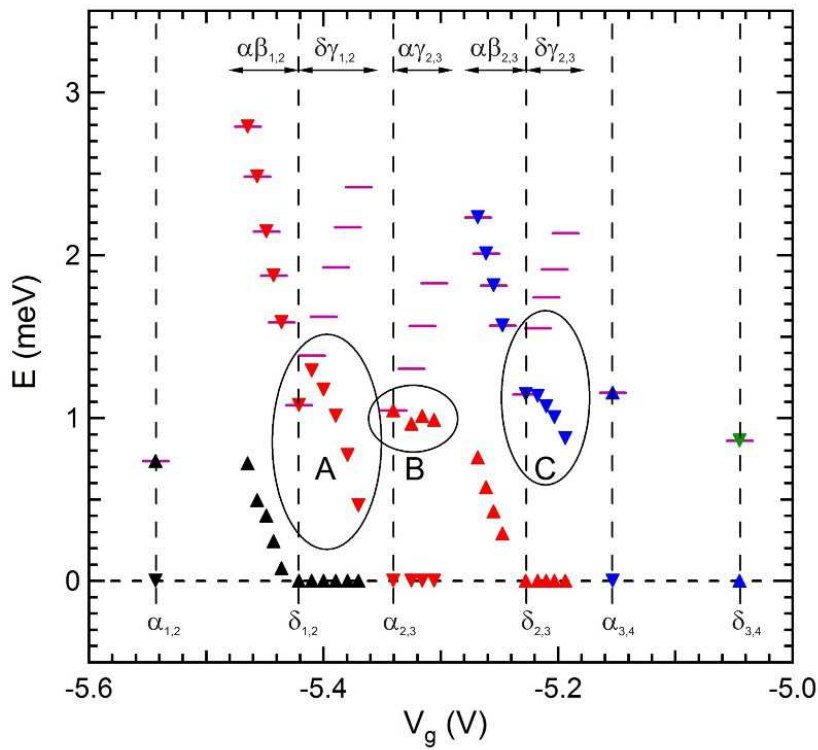}
\caption{The gate-voltage $V_{g}$ dependence of $\mu_{s}$
(horizontal purple lines) and various 1D subband edges (triangles).
Up (down) triangles correspond to spin-up (down) subbands and the
colours black, red, blue and green correspond to subband indices $n
= 1, 2, 3$ and $4$. The drain reservoir $\mu_{d}$ has been set as
the zero of energy. The corresponding configurations
$\alpha_{n,n+1}$, $\alpha\beta_{n,n+1}$, $\delta_{n,n+1}$
$\delta\gamma_{n,n+1}$ and $\alpha\gamma_{n,n+1}$ are indicated to
facilitate reference to Fig.~1. The circled regions A, B and C
indicate the behaviours of the $2\downarrow$, $2\uparrow$ and
$3\downarrow$ subbands as they populate, and are discussed in detail
in the text.}
\end{figure}

To demonstrate that direct tracking of the 1D subband edges is
possible, we have reanalysed the data presented in
Ref.~\cite{ChenPRB09}. This was achieved by using the {\it
Datathief} software package~\cite{datathief} to extract the $g^{*}$
versus $V_{g}$ data in Fig.~4 of Ref.~\cite{ChenPRB09}. From
Figs.~3(a,b) of Ref.~\cite{ChenPRB09} we similarly extracted the
subband spacings $\Delta E_{n\downarrow,n+1\downarrow}$ and $\Delta
E_{n\uparrow,n+1\uparrow}$, and the source-drain bias values
$V_{sd}$ versus $V_{g}$ for each $g^{*}$ value shown in Fig.~4 of
Ref.~\cite{ChenPRB09}. This made it possible to work backwards for
each point to find, relative to $\mu_{d}$, the energies of
$\mu_{s}$, the subband edge held at a reservoir potential and the
subband edge in the bias window. These are plotted at their
corresponding $V_{g}$ values in Fig.~3. This figure is complex at
first sight, we will therefore explain it step-by-step. The drain is
at electrical ground (via the current input of the lock-in
amplifier) in this measurement configuration, and hence we consider
$\mu_{d}$ as our zero of energy, in keeping with convention in
related papers on QPCs. The drain potential is denoted by the black
dashed horizontal line in Fig.~3. The horizontal purple bars show
the position of $\mu_{s}$ relative to $\mu_{d}$ at each $V_{g}$,
which is obtained from the corresponding point on the zig-zag path
shown in Fig.~1(a), when applied to Fig.~3(a) of
Ref.~\cite{ChenPRB09}. The locations of the subband edges that are
within the dc bias window are shown as solid symbols~\footnote{For
completeness, in Supplementary Fig.~1, we present a copy of Fig.~3
with an additional series of hollow data points that follow the same
shape/colour convention as those presented in Fig.~3. These points
correspond to subband edges that nominally fall outside the dc bias
window. Their locations are estimated by assuming that $\Delta
E_{n\downarrow,n+1\downarrow}$ and $\Delta
E_{n\uparrow,n+1\uparrow}$ are $V_{g}$ independent.}, with upward
(downward) triangles indicating spin up (down), and the colours
black, red, blue and green indicating $n = 1,2,3$ and $4$,
respectively, to best match the colour scheme used in Fig.~4 of
Ref.~\cite{ChenPRB09}.

We are limited to the data available in Ref.~\cite{ChenPRB09},
however it is possible to obtain a complete picture of both spin up
and spin down components for the second subband, as well as the spin
down component of the third subband. The relevant data are circled
and marked A, B and C, and lead to some interesting conclusions. In
Region A, we start at $\delta_{1,2}$, where the $2\downarrow$ edge
coincides with $\mu_{s}$ and the $1\uparrow$ edge coincides with
$\mu_{d}$, and follow the $\delta\gamma_{1,2}$ branch where the
$1\uparrow$ edge is held at $\mu_{d}$ and $\mu_{s}$ is gradually
raised in energy (see Fig.~1(f)). As the $2\downarrow$ subband
populates, it drops in energy, consistent with previous experimental
findings by Graham {\it et al}~\cite{GrahamPRB05}. Similar behaviour
is observed for the $3\downarrow$ subband in Region C.
Interestingly, $2\downarrow$ appears to briefly track $\mu_{s}$ as
$V_{sd}$ is initially increased. This initial delay in population is
also apparent for $3\downarrow$, albeit to a lesser extent. Delayed
population of $\downarrow$ subbands has not been previously reported
to our knowledge, and would be worth further investigation in future
studies.

Turning to Region B, here we start at $\alpha_{2,3}$, where the
$2\uparrow$ edge coincides with $\mu_{s}$ and the $2\downarrow$ edge
coincides with $\mu_{d}$, and follow the $\alpha\gamma_{2,3}$ branch
where the $2\downarrow$ edge is held at $\mu_{d}$, and $\mu_{s}$ is
gradually raised in energy (see Fig.~1(c)). Here it is clear that
$2\uparrow$ initially populates more slowly than $2\downarrow$, as
pointed out by Chen {\it et al}~\cite{ChenPRB09}, and in general
agreement with studies by Graham {\it et al}~\cite{GrahamPRB07}. A
question of interest at present is whether the $\uparrow$ subbands
pin to $\mu_{s}$ as they populate~\cite{GrahamPRB07, LasslPRB07} or
merely appear to due to a relatively slow population (see e.g.,
Fig.~1(b) of Ref.~\cite{ReillyPhysE06}, or Ref.~\cite{LindPRB11}).
Our re-analysis in Fig.~3 shows no evidence that the $2\uparrow$
edge pins to $\mu_{s}$ after the $2\downarrow$ subband edge reaches
$\mu_{d}$, and due to the nature of the method and limited available
data, it is not possible to accurately comment on the behaviour of
$2\uparrow$ before $2\downarrow$ reaches $\mu_{d}$ at $V_{g} =
-5.34$~V ($\alpha_{2,3}$ in Fig.~3). Note that although there is a
missing branch between the $\delta\gamma_{1,2}$ branch corresponding
to region A and the $\alpha\gamma_{2,3}$ branch corresponding to
region B in Fig.~3, it would provide no new information about the
behaviour of $2\uparrow$ relative to $\mu_{s}$. This is because
$2\uparrow$ is held at $\mu_{s}$ throughout this $\delta\beta_{1,2}$
branch. However, it does provide new information about the motion of
$2\uparrow$ relative to $2\downarrow$ that may be useful. By
following other paths in the $V_{sd}$ versus $V_{g}$ space (e.g.,
vertical motion by changing $V_{sd}$ at specifically chosen $V_{g}$)
it may be possible to extract additional knowledge about subband
edge motion and pinning without reliance on assumptions about
subband spacing.

Another question of interest is how the gap between $\uparrow$ and
$\downarrow$ subbands evolves after the $\uparrow$ subband has
dropped below $\mu_{s}$ -- does the gap keep
opening~\cite{ReillyPRL02}, hold constant (see Fig.~5 of
Ref.~\cite{ChenPRB09}, or start closing again~\cite{WangPRB96,
LasslPRB07, LindPRB11}? The $\alpha\beta$ branch data between
Regions B and C suggests that as $2\uparrow$ approaches $\mu_{d}$ it
populates at a very similar rate to the $2\downarrow$ subband under
similar circumstances. Very similar behaviour is observed for
$1\uparrow$, and is consistent with recent measurements by Chen {\it
et al}~\cite{ChenNL10}. There a study of plateau-like structures at
$0.7 - 0.85 G_{0}$ in $G_{ac}$ and $G_{dc}$ at finite $V_{sd}$ led
Chen {\it et al} to conclude that there is an unusual population
behaviour for the first spin-up subband as it moves between
$\mu_{s}$ and $\mu_{d}$. The more rapid drop in the $\uparrow$
subbands as $V_{g}$ becomes less negative, observed in Fig.~3,
suggests at least a stabilization of the spin-gap (i.e., it becomes
constant in $V_{g}$), and perhaps that it may even close again.
However, this latter possibility depends on how $2\downarrow$ moves
once it is below $\mu_{d}$, something that is inaccessible to these
measurements~\footnote{Note that at first this seems at odds with
the data in Fig.~4 of Ref.~\cite{ChenPRB09} where $g^{*}_{2}$ keeps
increasing monotonically, however this is based on the assumption of
constant $\Delta E_{n\downarrow,n+1\downarrow}$, this is clear by
examining Supplementary Fig.~1.}. This behaviour, if it occurs,
would be consistent with the Bruus, Cheianov and Flensberg
model~\cite{BruusPhysE01,KristensenPS02} and recent calculations by
both Jaksch {\it et al}~\cite{JakschPRB06} and Lind {\it et
al}~\cite{LindPRB11}. Clearly further measurements using this
approach are warranted to look more closely at the evolution of the
subbands as they are populated. It would be particularly interesting
to use a device where independent control over the QPC width and the
2DEG density could be achieved, for example, by a top- or back-gate
as in the devices studied by Reilly {\it et al}~\cite{ReillyPRL02}
or Hamilton {\it et al}~\cite{HamiltonAPL92}, respectively.

\section{Conclusions and Outlook}

In conclusion, we have provided a general framework for extracting
1D subband edge energies and $g$-factors using ac and dc conductance
measurements of QPCs. This framework shows routes to improving the
accuracy of measured $g^{*}$ values and interesting opportunities
for additional measurements, in particular, tracking of the 2nd
subband $g$-factor over the population of the $2\downarrow$ subband.
It also demonstrates that the measured data do not exhibit trends
consistent with calculations by Wang and Berggren~\cite{WangPRB96}
but may instead point to a density-dependent spin-gap as predicted
by Reilly {\it et al}~\cite{ReillyPRL02, ReillyPRB05,
ReillyPhysE06}. Finally, we show that the information extracted from
dc conductance measurements can be used to map the evolution of the
1D subband edges with $V_{g}$ and may provide more useful knowledge
about the physics occurring as the 1D subbands populate than
conversion to a $g$-factor does. In particular, an analysis from a
subband energy perspective shows that the $2\downarrow$ subband
drops in energy as it populates, consistent with earlier
measurements by Graham {\it et al}~\cite{GrahamPRB05}, but suggests
that the $2\downarrow$ edge subband tracks $\mu_{s}$ closely at
first, a feature not previously reported in the literature. The
$2\uparrow$ subband initially populates more slowly, in general
agreement with earlier work by Graham {\it et
al}~\cite{GrahamPRB07}. There is no evidence that the $2\uparrow$
edge pins to $\mu_{s}$, however it is not possible to measure this
until $2\downarrow$ reaches $\mu_{d}$ using the data available in
Ref.~\cite{ChenPRB09}. Our re-analysis also shows that the
population rate for $2\uparrow$ eventually increases to become as
rapid as that for $2\downarrow$. This suggests that the spin-gap may
become independent of $V_{g}$, and perhaps even close again, as the
$2\uparrow$ subband continues to populate. This behaviour would be
in rough qualitative agreement with theoretical calculations
~\cite{WangPRB96, JakschPRB06, LasslPRB07, LindPRB11}, and is
consistent with the suggestion by Chen {\it et al}~\cite{ChenNL10}
that there is an unusual population behaviour of the first spin-up
subband as it passes between the source and drain potentials. Our
re-analysis highlights the opportunity for further measurements with
this approach, particularly in devices where the QPC width and
electron density can be tuned independently.

\ack APM acknowledges financial support from an Australian Research
Council (ARC) Future Fellowship (FT0990285), and the ARC Discovery
Projects (DP0877208 and DP110103802) and Linkage International
(LX0882222) schemes. We thank T.P. Martin and A.R. Hamilton for
several helpful discussions on the physics, R. Newbury and S. Fricke
for proof-reading of the manuscript, and T.-M. Chen and A.R.
Hamilton for helpful conversations regarding experimental technique
for these measurements.


\section{References}

\begin{thebibliography}:

\bibitem{BerggrenPW02} K.-F. Berggren and M. Pepper, Physics World {\bf 15(10)}, 37 (2002).

\bibitem{ChoAPL71} A.Y. Cho, Appl. Phys. Lett. {\bf 19}, 467 (1971).

\bibitem{ThorntonPRL86} T.J. Thornton, M. Pepper, H. Ahmed, D. Andrews and G.J. Davies, Phys. Rev. Lett. {\bf 56}, 1198 (1986).

\bibitem{vanWeesPRL88} B.J. van Wees, H. van Houten, C.W.J. Beenakker, J.G. Williamson, L.P. Kouwenhoven, D. van der Marel and C.T. Foxon, Phys. Rev. Lett. {\bf 60}, 848 (1988).

\bibitem{WharamJPC88} D.A. Wharam, T.J. Thornton, R. Newbury, M. Pepper, H. Ahmed, J.E.F. Frost, D.G. Hasko, D.C. Peacock, D.A. Ritchie and G.A.C. Jones, J. Phys. C {\bf 21}, L209 (1988).

\bibitem{ThomasPRL96} K.J. Thomas, J.T. Nicholls, M.Y. Simmons, M. Pepper, D.R. Mace and D.A. Ritchie, Phys. Rev. Lett. {\bf 77}, 135 (1996).

\bibitem{WangPRB98} C.-K. Wang and K.-F. Berggren, Phys. Rev. B {\bf 57}, 4552 (1998).

\bibitem{MeirPRL02} Y. Meir, K. Hirose and N.S. Wingreen, Phys. Rev. Lett. {\bf 89}, 196802 (2002).

\bibitem{CronenwettPRL02} S.M. Cronenwett, H.J. Lynch, D. Goldhaber-Gordon, L.P. Kouwenhoven, C.M. Marcus, K. Hirose, N.S. Wingreen and V. Umansky, Phys. Rev. Lett {\bf 88}, 226805 (2002).

\bibitem{KristensenPRB00} A. Kristensen, H. Bruus, A.E. Hansen, J.B. Jensen, P.E. Lindelof, C.J. Marckmann, J. Nyg{\aa}rd, C.B. S{\o}renson, F. Beuscher A. Forchel and M. Michel, Phys. Rev. B {\bf 62}, 10950 (2000).

\bibitem{BruusPhysE01} H. Bruus, V.V. Cheianov and K. Flensberg, Physica E {\bf 10}, 97 (2001).

\bibitem{ReillyPRL02} D.J. Reilly, T.M. Buehler, J.L. O'Brien, A.R. Hamilton, A.S. Dzurak, R.G. Clark, B.E. Kane, L.N. Pfeiffer and K.W. West, Phys. Rev. Lett. {\bf 89}, 246801 (2002).

\bibitem{ReillyPRB05} D.J. Reilly, Phys. Rev. B {\bf 72}, 033309 (2005).

\bibitem{ReillyPhysE06} D.J. Reilly, Y. Zhang and L. DiCarlo, Physica E {\bf 34}, 27 (2006).

\bibitem{GrahamPRB05} A.C. Graham, M. Pepper, M.Y. Simmons and D.A. Ritchie, Phys. Rev. B {\bf 72}, 193305 (2005).

\bibitem{GrahamPRB07} A.C. Graham, D.L. Sawkey, M. Pepper, M.Y. Simmons and D.A. Ritchie, Phys. Rev. B {\bf 75}, 035331 (2007).

\bibitem{ChenAPL08} T.-M. Chen, A.C. Graham, M. Pepper, I. Farrer and D.A. Ritchie, Appl. Phys. Lett. {\bf 93}, 032102 (2008).

\bibitem{ChenPRB09} T.-M. Chen, A.C. Graham, M. Pepper, F. Sfigakis, I. Farrer and D.A. Ritchie, Phys. Rev. B {\bf 79}, 081301 (2009).

\bibitem{ChenNL10} T.-M. Chen, A.C. Graham, M. Pepper, I. Farrer, D. Anderson, G.A.C. Jones and D.A. Ritchie, Nano Lett. {\bf 10}, 2330 (2010).

\bibitem{PatelPRB91} N.K. Patel, J.T. Nicholls, L. Martin-Moreno, M. Pepper, J.E.F. Frost, D.A. Ritchie and G.A.C. Jones, Phys. Rev. B {\bf 44}, 10973 (1991).

\bibitem{DaneshvarPRB97} A.J. Daneshvar, C.J.B. Ford, A.R. Hamilton, M.Y. Simmons, M. Pepper and D.A. Ritchie, Phys. Rev. B {\bf 55}, 13409 (1997).

\bibitem{DanneauPRL06} R. Danneau, O. Klochan, W.R. Clarke, L.H. Ho, A.P. Micolich, M.Y. Simmons, A.R. Hamilton, M. Pepper, D.A. Ritchie and U. Z\"{u}licke, Phys. Rev. Lett. {\bf 97}, 026408 (2006).

\bibitem{SchapersAPL07} Th. Sch\"{a}pers, V.A. Guzenko and H. Hardtdegen, Appl. Phys. Lett. {\bf 90}, 122107 (2007).

\bibitem{MartinAPL08} T.P. Martin, A. Szorkovszky, A.P. Micolich, A.R. Hamilton, C.A. Marlow, H. Linke, R.P. Taylor and L. Samuelson, Appl. Phys. Lett. {\bf 93}, 012105 (2008).

\bibitem{MartinPRB10} T.P. Martin, A. Szorkovszky, A.P. Micolich, A.R. Hamilton, C.A. Marlow, R.P. Taylor, H. Linke and H.Q. Xu, Phys. Rev. B {\bf 81}, 041303 (2010).

\bibitem{PatelPRB91A} N.K. Patel, J.T. Nicholls, L. Martin-Moreno, M. Pepper, J.E.F. Frost, D.A. Ritchie and G.A.C. Jones, Phys. Rev. B {\bf 44}, 13549 (1991).

\bibitem{pc} A.R. Hamilton and T.-M. Chen (Private Communication).

\bibitem{ButtikerPRB90} M. B\"{u}ttiker, Phys. Rev. B {\bf 41}, 7906 (1990).

\bibitem{AndoJPSJ74} T. Ando and Y. Uemura, J. Phys. Soc. Jpn {\bf 37}, 1044 (1974).

\bibitem{WangPRB96} C.-K. Wang and K.-F. Berggren, Phys. Rev. B {\bf 54}, 14257 (1996).

\bibitem{datathief} B. Tummers, J. van der Lann and K. Huyser, {\it DataThief III V1.6}, Available at http://www.datathief.org.

\bibitem{KristensenPS02} A. Kristensen and H. Bruus, Phys. Scripta {\bf T101}, 151 (2002).

\bibitem{JakschPRB06} P. Jaksch, I. Yakimenko and K.-F. Berggren, Phys. Rev. B {\bf 74}, 235320 (2006).

\bibitem{LasslPRB07} A. Lassl, P. Schlagheck and K. Richter, Phys. Rev. B {\bf 75}, 045346 (2007).

\bibitem{LindPRB11} H. Lind, I. Yakimenko and K.-F. Berggren, Phys. Rev. B {\bf 83}, 075308 (2011).

\bibitem{HamiltonAPL92} A.R. Hamilton, J.E.F. Frost, C.G. Smith, M.J. Kelly, E.H. Linfield, C.J.B. Ford, D.A. Ritchie, G.A.C. Jones, M. Pepper, D.G. Hasko and H. Ahmed, Appl. Phys. Lett. {\bf 60}, 2782 (1992).

\end{thebibliography}
\end{document}